\documentclass[onecolumn,preprint,amsmath,amssymb,aps,prl]{revtex4}

\usepackage{graphicx}
\usepackage{dcolumn}
\usepackage{bm}
\usepackage{textcomp}

\begin{document}

\preprint{APS/123-QED}

\title{Phase Conjugation and Negative Refraction Using Nonlinear Active
Metamaterials}

\author{Alexander R. Katko$^1$}
\author{Shi Gu$^1$}
\author{John P. Barrett$^1$}
\author{Bogdan-Ioan Popa$^1$}
\author{Gennady Shvets$^2$}
\author{Steven A. Cummer$^1$}\email{cummer@ee.duke.edu}

\affiliation{$^1$Department of Electrical and Computer Engineering and Center for Metamaterials and Integrated Plasmonics, Duke
University, Durham NC 27708\\
$^2$Department of Physics, The University of Texas at
Austin, Austin TX 78712}

\date{\today}

\begin{abstract}
We present experimental demonstration of phase conjugation using
nonlinear metamaterial elements. Active split-ring resonators
loaded with varactor diodes are demonstrated theoretically to act
as phase-conjugating or time-reversing discrete elements when
parametrically pumped and illuminated with appropriate frequencies. The metamaterial elements were fabricated
and shown experimentally to produce a time reversed signal.
Measurements confirm that a discrete array of phase-conjugating
elements act as a negatively-refracting time reversal RF lens
only 0.12$\lambda$ thick.  
\end{abstract}


\maketitle

Metamaterials are subwavelength structures which, when viewed
macroscopically, achieve electromagnetic properties different from
their constituent elements. This may include an effective negative
refractive index \cite{smith}, negative effective permeability
\cite{pendrySRR}, effective permittivity near zero
\cite{ENZpaper}, or many other unusual properties. Equally importantly, functionality can be embedded into metamaterials that can be difficult if not impossible to find in conventional materials. Examples of this approach include active metamaterials
with~\cite{zlossactive,bogdanactive,fang_PRBcalc} and
without~\cite{padilla_nature06,reynet_apl,wang_apl07} gain for loss compensation and
frequency tuning, respectively, as well as amplification \cite{Syms_metamat2008}. Embedding nonlinear response into metamaterials has also been used for power-dependent tuning~\cite{costas_optexp08,smithnonlinear,nonlinearSRRs}. Coherent harmonic generation has also been demonstrated across the
electromagnetic spectrum~\cite{activenonlinear,wegener_science06}
using nonlinear metamaterials.

In this Letter we introduce a new class of nonlinear active
metamaterials (NAMMs). Combining metamaterials' nonlinear response
with an active high-frequency external excitation in the simplest
metamaterial building block, the split-ring resonator (SRR), we
experimentally demonstrate phase conjugation of microwave signals. Using
numerical simulations, we demonstrate that such NAMM can be used
as a building block for a negatively-refracting lens. Negative
refraction~\cite{veselago} is one of the most discussed and
intriguing phenomena exhibited by a special class of metamaterials
with negative permittivity and permeability. It was shown
theoretically by Pendry \cite{PendryTimereversal}, however, that
negative refraction could also be produced by a thin slab of a
material exhibiting time reversal (or, equivalently, phase
conjugation). Traditional thick slabs of phase-conjugating (PC)
media have been used for the useful property of generating
negative frequency waves that focus back on the source, sometimes
called retrodirectivity \cite{PCYariv}. Discrete arrays of PC elements have also been used for retrodirectivity by the microwave community \cite{ChiuRetroarray,PonRetroarray}. Pendry showed that a thin slab would indeed generate these retrodirective waves but would also generate a wave on the other side of the material exhibiting negative refraction. A few recent works have examined using discrete PC elements for forward focusing \cite{subwaveExp,IEEEphaseconj}, but these cannot be extended to create an effective medium. As an example, \cite{subwaveExp} used an array of patch antennas for receiving; a phase-conjugating circuit for each antenna composed of PLLs, filters, phase shifters, IQ modulators, and other components; and a separate array of patch antennas for retransmission. Using NAMMs, we demonstrate negative refraction using simple metamaterial elements with varactors embedded, allowing easy extension to a true volumetric approach.

We demonstrate theoretically and experimentally
that nonlinear metamaterial elements, when driven with an active
source, act as PC elements. The nonlinear
characteristics of the metamaterials are obtained by embedding
varactor diodes in each metamaterial element and exciting each
element with a monochromatic RF signal, allowing easy extension to
construction of a bulk medium of phase-conjugating SRRs with no
external circuitry. We then demonstrate experimentally that
nonlinear metamaterials do indeed act as phase-conjugating and
thus negatively-refracting elements, and in an array they act as a time-reversing material.

Parametric resonance can be used to realize phase-conjugation in a single cell of a
NAMM. Consider a single dipole-like nonlinear resonant scatterer
with the parameterically modulated natural frequency $\Omega(t)$
given by $\Omega^2(t) = \Omega_0^2 \left( 1 + h_0
\sin{(2\omega_0t)} \right)$, where the $\Omega(t) =
1/\sqrt{L(t)C(t)}$ is related to the resonator's inductance $L$
and capacitance $C$, one (or both) of which are modulated with the
frequency $2\omega_0$ by the externally provided signal. The dimensionless parameter $h_0$ characterizes the strength of the parametric drive. Our experimental implementation relies on the modulation of the SRR's capacitance by inserting a nonlinear varactor whose capacitance depends on the externally-applied high-frequency voltage. Further
assume that SRRs are arrayed in the $z=0$ plane along the $y-$
axis and subjected to an external monochromatic electromagnetic wave $E(y,t)=\sum_{k_y} E_{k_y} \exp{[-i(\omega_1 t - k_y y)]} +
c.~c.$ incident from the negative $z$ direction and produced by a far-field source located at $z_{\rm source} < 0$. The response of a
single $j$'s SRR, characterized by its induced dipole moment
$p_j$, is calculated according to
\begin{eqnarray}\label{eq:dipole_moment}
  &&\frac{d^2 p_j}{dt^2} + \gamma \frac{dp_j}{dt} + \Omega_0^2 \left( 1 + h_0 \sin{(2\omega_0t)} \right) p_j = \nonumber \\ &&\sum_{k_y} \Omega_0^2 E(k_y) a^3 e^{-i(\omega_1 t - k_y y_j)} + c.c.,
\end{eqnarray}
where $y_j$ is the spatial location of the SRR-based NAMM, $a$ is its effective size, and $\gamma$ is the SRR's loss rate that includes dissipative and radiative losses. The effects of inter-element interaction are assumed to be negligible owing to the tight field confinement. Eq.~(\ref{eq:dipole_moment}) models a single SRR as a linear electric dipole with the low-frequency polarizability $a^3$ which is resonantly-driven by the RF field of the antenna source. The solution of Eq.~(\ref{eq:dipole_moment}) can be expressed as $p_j = a_1^{(j)}\exp{(-i\omega_1 t)} + b_2^{(j)} \exp{(i\omega_2 t)}$, where $\omega_2 = 2\omega_0 - \omega_1$, and $a_1^{(j)},b_2^{(j)}$ are computed by substituting $p_j$ into Eq.~(\ref{eq:dipole_moment}) to obtain:

\begin{equation}\label{eq:a1b2}
    a_1^{(j)} = \frac{\Omega^2-\omega_2^2+2i\gamma
    \omega_2}{D(\omega_1)} e^{ik_y y_j}, b_2^{(j)} = \frac{i h_0 \Omega^2}{2D(\omega_1)} e^{ik_y y_j},
\end{equation}
where the dispersion function $D(\omega_1)$ is given by
\begin{equation}\label{eq:determinant}
    D = (\Omega^2 - \omega_1^2 -2i\gamma \omega_1) (\Omega^2 - \omega_2^2 + 2i\gamma \omega_2) - h_0^2 \Omega^4/4.
\end{equation}
The electric field at $z>0$ behind the plane of NAMMs is given by the sum of the incident fields and the fields produced by the induced dipole moments $p_j$ of all NAMMs. Because the components of $p_j$ proportional to $a_1$ have the same spatio-temporal dependence as the incident electric field $E(y,t)$, they produce scattered radiation which is spatially diverging away from the source. However, the contributions to $p_j$  proportional to $b_2$ produce a phase-conjugated signal which, assuming that $\omega_2 \approx \omega_1$, produces a spatially-converging field with the focus at $z_{\rm image} = - z_{\rm source} > 0$, i.~e.~the plane of NAMMs constitutes a flat negative-refraction lens. 

The strongest phase-conjugated signal is produced when $\Omega_0 = \omega_0$ and $\omega_1 \approx \omega_0$. However, $\omega_2$ can deviate from $\omega_1$, significantly simplifying our experiments because the phase-conjugated signal has a different frequency from that of the incident field and is not overwhelmed by it. Note that, for certain values of $\omega_0$, $\Omega_0$, $\gamma$, and $h_0$, there exist complex solutions $\omega_{\ast} \equiv \omega_{\ast}^{(r)} + i\omega_{\ast}^{(i)}$ of $D(\omega_{\ast})=0$ corresponding to
the parametric instability~\cite{landau_mech} with the growth rate $\omega_{\ast}^{(i)}$. For example, for $\omega=\Omega_0$ instability occurs if $h_0 \omega_0 > 2\gamma_0$. For the rest of this Letter we assume that the strength of the parametric drive $h_0$ is small enough for the NAMM to remain stable.

Following the above analysis, we designed and fabricated SRR-varactor nonlinear metamaterial elements. They were designed to resonate near 950 MHz and have a pump signal delivered via cables directly to the varactors.  Fig. 1 shows the layout of an SRR with dimensions and corresponding photo of a fabricated SRR.

\begin{figure}[h!]
\includegraphics[width=3in]{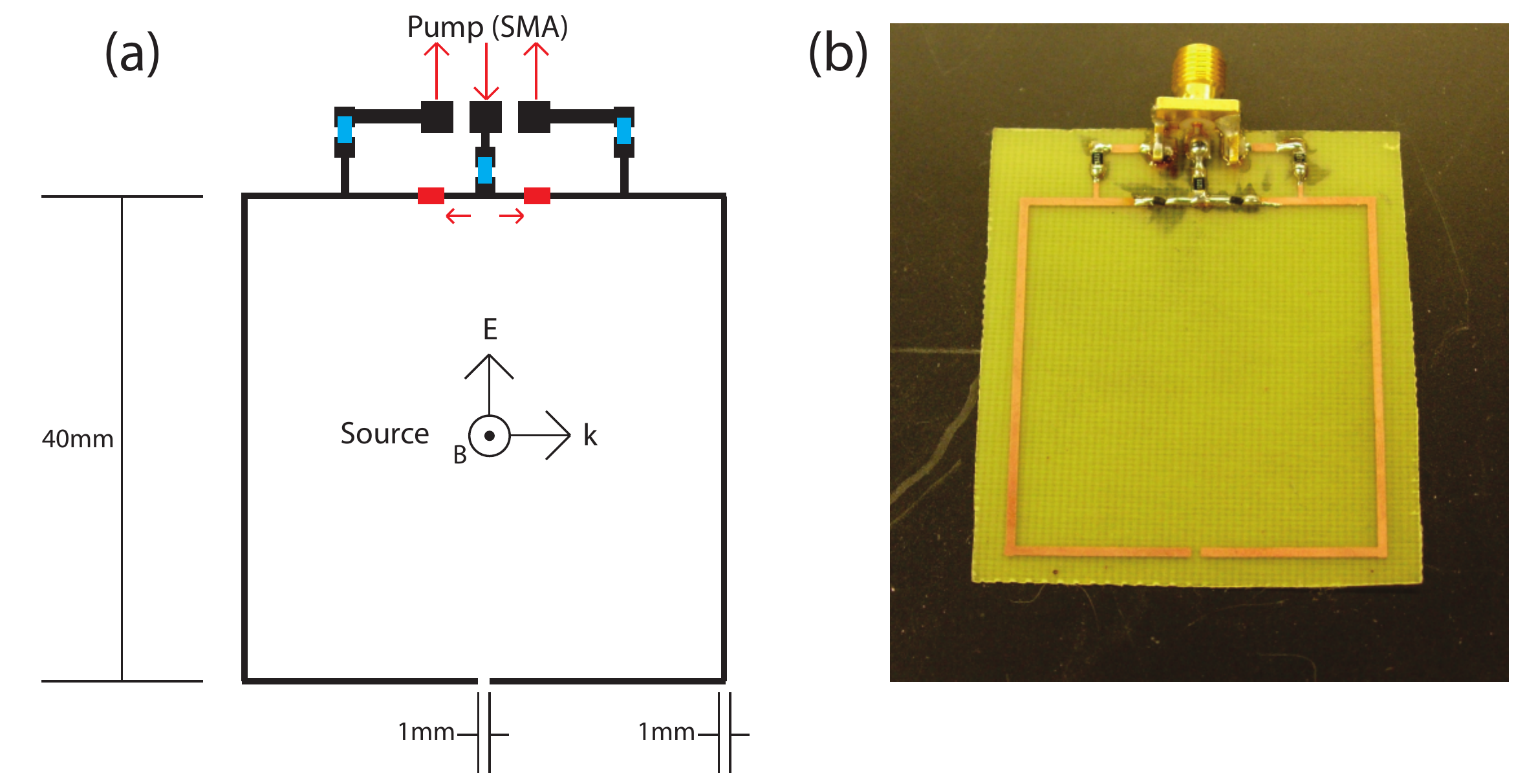}
\vfill \caption{(Color online) (a): Layout of SRR element. Red
elements are varactor diodes and blue elements are isolation resistors. (b):
Photo of fabricated SRR element. }
\end{figure}

The SRRs were pumped with a signal at a specified
pump frequency $f_{pump}$ corresponding to $2\omega_0$. A monopole antenna located above a ground plane was used as the illumination source, radiating a signal at a specified source frequency $f_{source}$ corresponding to $\omega_1$.
As described above, a nonlinearly generated, phase conjugated signal should be reradiated at $f_{pc}=f_{pump}-f_{source}$ corresponding to $\omega_2$.

Measurements were made using a spectrum analyzer with a receiving
antenna to determine interferene-free frequency bands and the frequencies were
selected so that the mixed signal at $f_{pc}$ was clearly
distinct from the source signal at $f_{source}$. These experiments
used frequencies of $f_{pump}=1850$MHz, $f_{source}=950$MHz, and
subsequently $f_{pc}=900$MHz. Output from the spectrum analyzer
is shown in Fig. 2 showing both the background signal levels and
signal levels for the experiments with the pump and source signal
generators switched on. As expected with our chosen frequencies,
the signals at $f_{pump}=1850$MHz, $f_{source}=950$MHz, and $f_{pc}=900$MHz are only due to our experimental setup and not background radiation, demonstrating that the nonlinear SRRs are indeed generating a nonlinear mixed signal. The maximum measured value of the signal at $f_{source}$ was -25.2 dBm. Through the experiments the maximum recorded value of the PC signal at $f_{pc}$ was -49.6 dBm, yielding a minimum ratio of -24.4 dB between the source and PC signal. The source of this difference comes from the mixing efficiency in our experiments, which was not optimized in this work. Improving the coupling efficiency of the pump signal into the nonlinear metamaterial to overcome losses in the SRR will result in more efficient and practically-relevant NAMMs. The signal at 925MHz corresponds to the sub-harmonic of $\frac{f_{pump}}{2}$ and is present due to a power amplifier used in connection with the pump. The broad signal from approximately 868 to 888MHz corresponds to the downlink band of GSM-850 cellular phone signals. This and the signal at 929MHz are independent of our experiments but did not interfere at any frequencies. No higher order mixed frequency signals were detected and, along with the noted weak PC signal at least 24.4 dB down from the source, the assumptions in the analysis section are shown to be valid.

\begin{figure}[h!]
\includegraphics[width=3in]{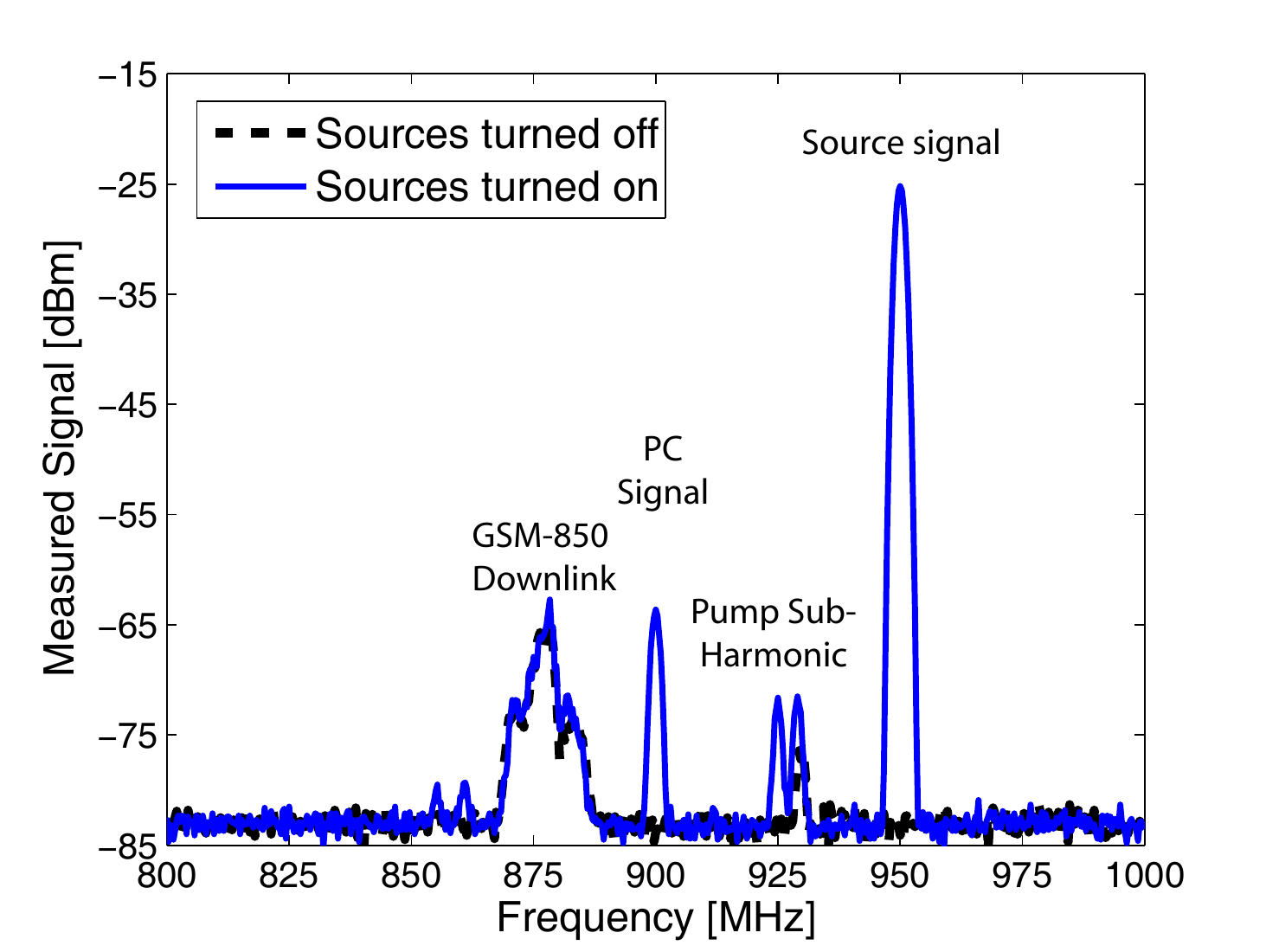}
\vfill \caption{(Color online) Data from spectrum analyzer for
background signal (dashed black) and pump and source signals
switched on (solid blue). Notable frequencies include 900MHz
($f_{pc}$), 950MHz ($f_{source}$), 1850MHz (not shown).}
\end{figure}

We conducted multiple experiments to demonstrate phase conjugation of these elements. To verify that individual elements were PC, we measured the interference pattern generated by two isolated elements. We also conducted an experiment to examine phase conjugation with an array of elements, described later. To map interference patterns, two monopole antennas above a ground plane were connected to a signal generator and spaced $1.5\lambda$ apart. A microwave absorber was placed between the antennas to isolate them. They were excited both in phase and --90\textdegree out of phase, and the interference patterns were mapped at a constant distance $\lambda$ from the line connecting the antennas, at the signal frequency, as illustrated in Fig. 3(b). For two normal, non-conjugating elements excited in phase and then --90\textdegree out of phase, the maximum of the pattern moved to the right. Then the same interference patterns were measured at the mixed frequency with the SRRs in place. The absorber was used to ensure that only one antenna was illuminating each SRR. Again the antennas were excited in phase and then --90\textdegree out of phase and the interference pattern was mapped along the same line as before, at the mixed frequency. The measurements were interpolated and a filter was applied in the spatial Fourier domain to remove noise. The results are shown in Fig. 3(a) and a schematic is shown in Fig. 3(b).

\begin{figure}[h!]
\includegraphics[width=3in]{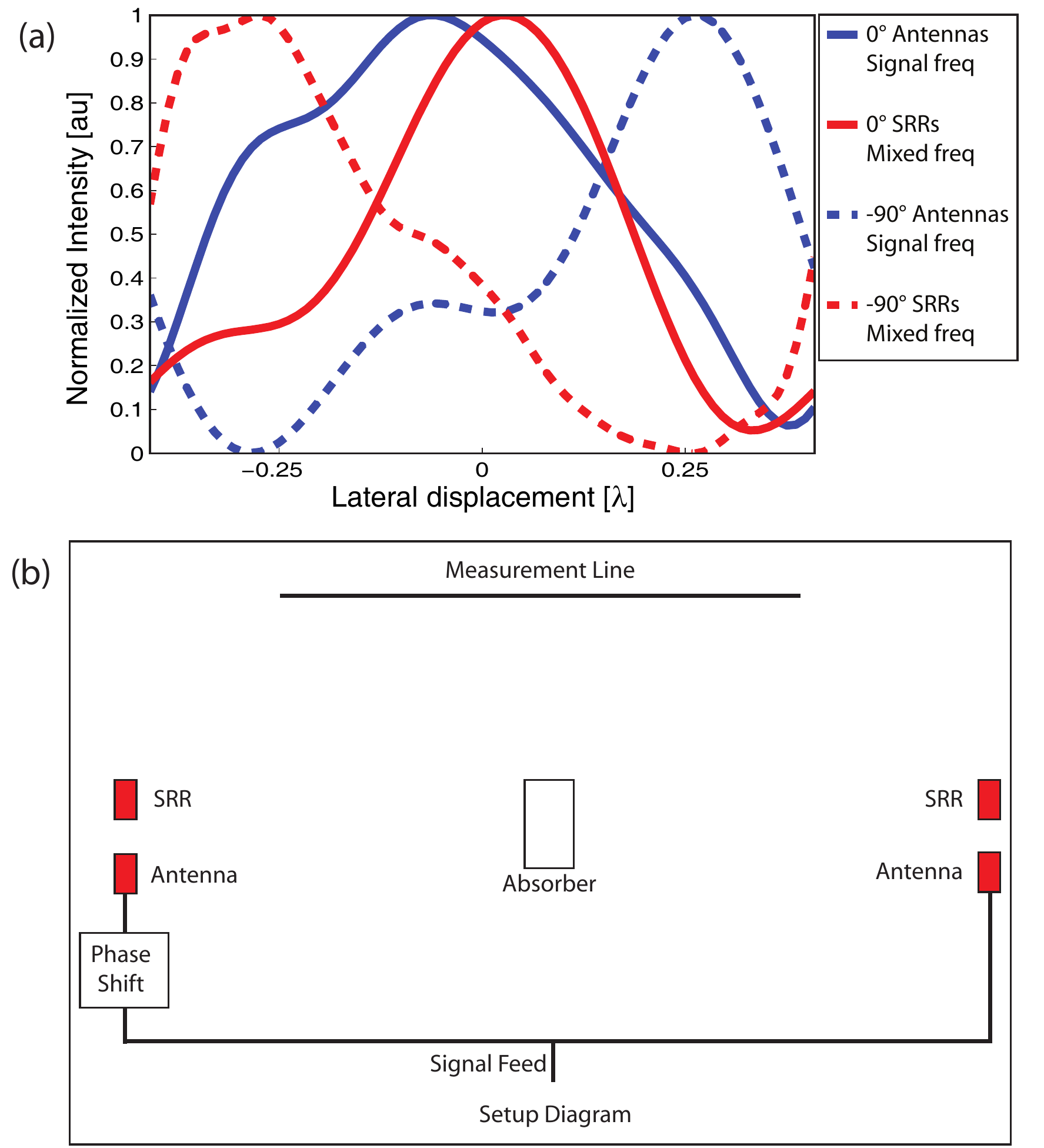}
\vfill \hfill \caption{(Color online) (a) Interference pattern for:
in-phase antennas at $f_{signal}$(solid blue), --90\textdegree out-of-phase
antennas at $f_{signal}$(dashed blue), in-phase SRRs at $f_{mixed}$(solid red), --90\textdegree out-of-phase SRRs at $f_{mixed}$(dashed red). The opposite lateral shifts in extrema between the SRRs and antennas, with the same phase shift applied to each, clearly demonstrates that the SRRs act as PC elements. (b) Schematic of setup (to scale). Antennas were fed at $f_{signal}$ with both 0\textdegree and --90\textdegree phase shifts. Absorber was used to isolate antennas and SRRs. The pump signal for the SRRs, as described above, is not shown.}
\end{figure}

Exciting two radiating elements with 0\textdegree phase shift and
the same spatial positioning yielded field patterns roughly independent of the elements used, as shown by the solid lines in Fig. 3. The small-scale features in each curve are artifacts of a low dynamic range in these measurements. However, the important feature is the clear difference in the locations of the global extrema. Introducing a phase shift to the non-conjugating elements creates a maximum to the right of the center. However, with the NAMM elements, the peak moves to the opposite side of the center, showing that exciting normal radiating elements with a --90\textdegree phase shift is equivalent to exciting the NAMM elements with a +90\textdegree phase shift.
This is a clear demonstration that the nonlinear SRRs act as phase-conjugating elements.

Three NAMM elements were then used to approximate a discrete slab of
time-reversal medium. The thickness of the SRR array was
approximately $0.12\lambda$. The SRRs were pumped in phase and illuminated with a single antenna. As before $f_{pump}=1850$MHz, $f_{source}=950$MHz, and
$f_{pc}=900$MHz. The SRRs were placed $\lambda/4$ apart,
creating an array a total of $\lambda/2$ in width. The source
antenna was placed a distance one $\lambda$ normal to the array and
$\lambda/2$ from the center of the array. The field power distribution of the mixed
and phase-conjugated signal was then measured spatially to create
a two-dimensional field map.

A simple calculation was used to generate the expected relative field distribution for phase-conjugating elements without considering details of the mixing efficiency of individual elements. The calculation assumed a $r^{-1}$ dependence on the original signal, the generation of the nonlinear difference frequency in linear proportion to the incident signal, and a $r^{-1}$ dependence and the measured radiation pattern of the SRRs for the reradiated signal. Its primary purpose was to predict the relative interference pattern for PC elements in a given geometry for comparison with measurements.
 
The field map was generated over a measurement area
of dimensions $-3\lambda/2$ to $3\lambda/2$ tangential to the
array and $\lambda/4$ to $\lambda$ normal to the array again by
using the spectrum analyzer.  The raw data was interpolated and a filter was applied in the spatial Fourier domain to remove noise from the measurements. Fig. 4 shows the results overlaid on the experimental setup.

\begin{figure}[h!]
\includegraphics[width=3in]{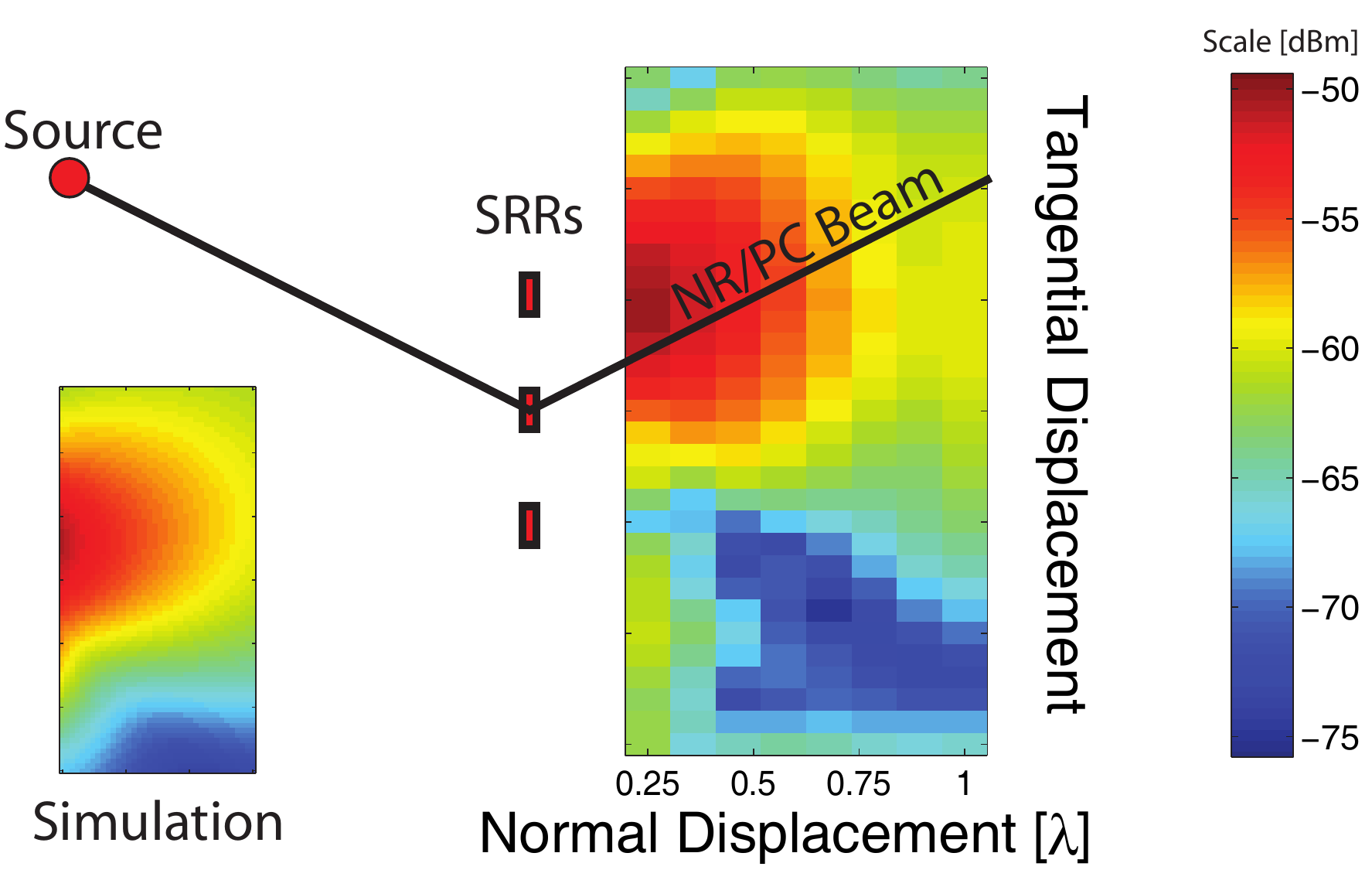}
\hfill \vfill \caption{(Color online) Measured $abs(E)_{dBm}$ and experimental setup. Setup is approximately to scale. The source radiates from the upper left in the figure and SRRs produce a mixed signal at $f_{pc}$. The inset shows calculated results, exhibiting good agreement. The measurements show that the beam negatively refracts.}
\end{figure}

The main beam at the phase-conjugated frequency $f_{pc}$ clearly refracts to the opposite side of the normal compared to a conventional material. Moreover, there is excellent agreement between measurement and calculation, as shown by the inset in the figure. The same calculation but assuming non-phase-conjugating elements was also conducted and yielded very different results. The field power along a line of constant distance from the array was examined to illustrate this more quantitatively,
comparing conjugate calculation, non-conjugate calculation, and filtered measured data. This is presented in Fig. 5.

\begin{figure}[h!]
\includegraphics[width=3in]{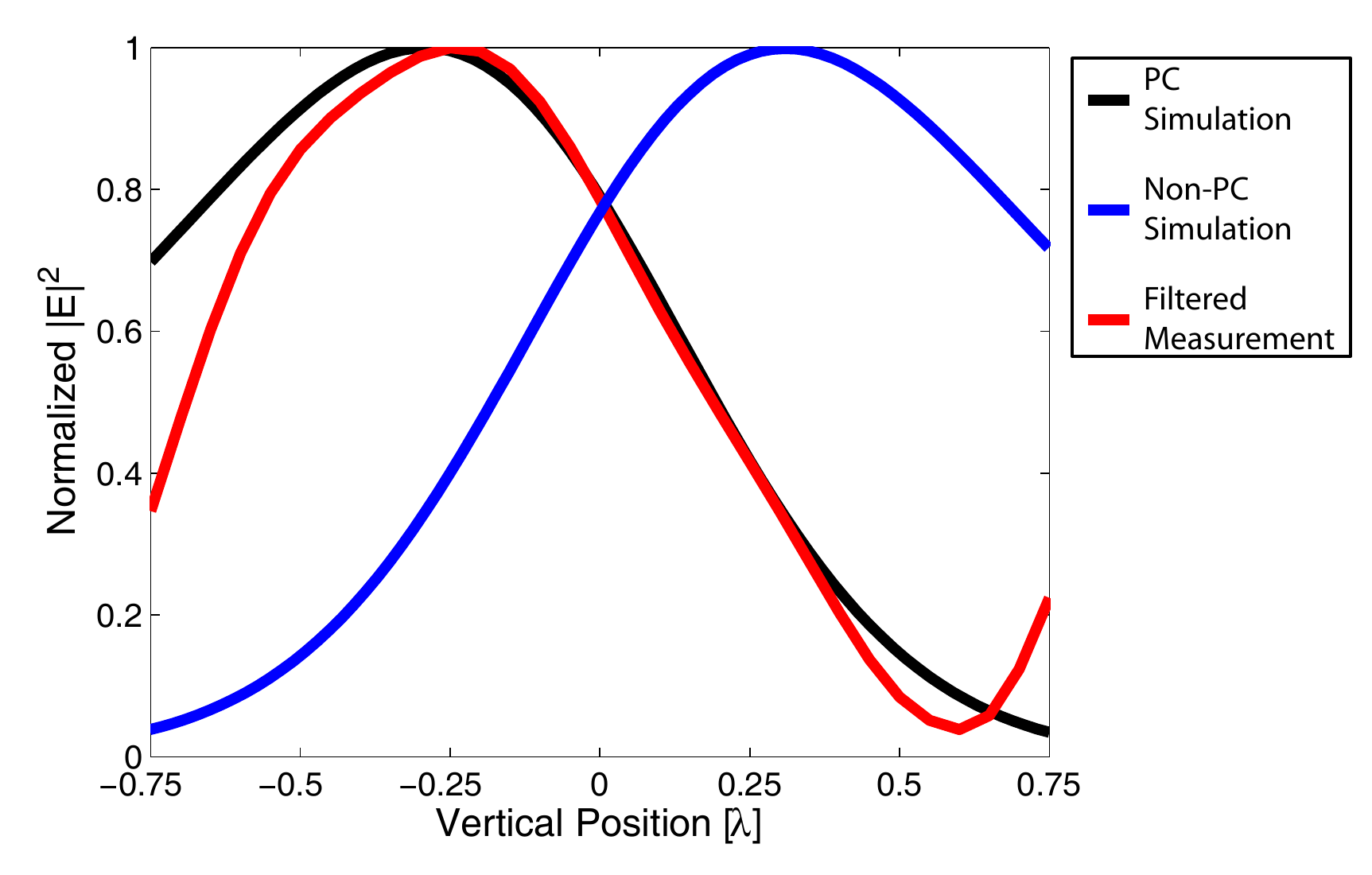}
\hfill \vfill \caption{(Color online) Normalized intensity at
constant distance from array for various data sets: calculated data
for phase-conjugating elements (black); calculated data for
non-phase-conjugating elements (blue); and filtered measured data (red). The measured data match calculated data very well and are negatively refracted when compared to non-PC simulated data.}
\end{figure}

An array of PC metamaterials is limited in far-field imaging ability only by normal imaging considerations like aperture size and spacing between elements. Creating a larger array with more PC elements provides a better approximation of an infinite, continuous medium and thus provides a closer approximation to an ideal far-field lens. The backward-focusing (retrodirectivity) of an array of PC elements has been explored to some extent \cite{ChiuRetroarray,PonRetroarray,IEEEphaseconj,subwaveExp}.
However, these approaches have relied on much more complex structures or circuitry and cannot be easily extended to volumetric devices. Our approach relies on much simpler elements which could be used as building blocks for PC effective media. Moreover, the primary application of retrodirective arrays has been for backward focusing. We have shown that NAMMs can be used to create a simple and effective phase-conjugating medium and provided a path towards realizing time-reversal metamaterial media.

In summary, we first demonstrated theoretically that metamaterials with
embedded nonlinear elements produce a phase-conjugated or time-reversed signal through parametric pumping. Two element experiments unambiguously demonstrated the phase-conjugation of the mixed signal, and measurements of the field distribution produced by a 3 element array showed the expected effective negative refraction behavior of the time-reversed signal.  Such elements form a simple but powerful building block for more complex phase conjugation media and devices, and we have shown that this class of active nonlinear metamaterials is promising for realising designs utilizing time-reversal media and thin-slab RF imaging.

Acknowledgements: this work was supported in part by DARPA under Contract No. HR0011-05-C-0068. G.S. would like to acknowledge the support of the U.S. Air Force Office of Scientific Research (AFOSR) MURI grants FA 9550-06-1-0279 and FA 9550-08-1-0394.


\begin{thebibliography}{22}
\expandafter\ifx\csname natexlab\endcsname\relax\def\natexlab#1{#1}\fi
\expandafter\ifx\csname bibnamefont\endcsname\relax
  \def\bibnamefont#1{#1}\fi
\expandafter\ifx\csname bibfnamefont\endcsname\relax
  \def\bibfnamefont#1{#1}\fi
\expandafter\ifx\csname citenamefont\endcsname\relax
  \def\citenamefont#1{#1}\fi
\expandafter\ifx\csname url\endcsname\relax
  \def\url#1{\texttt{#1}}\fi
\expandafter\ifx\csname urlprefix\endcsname\relax\def\urlprefix{URL }\fi
\providecommand{\bibinfo}[2]{#2}
\providecommand{\eprint}[2][]{\url{#2}}

\bibitem[{\citenamefont{Smith et~al.}(2000)\citenamefont{Smith, Padilla, Vier,
  Nemat-Nasser, and Schultz}}]{smith}
\bibinfo{author}{\bibfnamefont{D.~R.} \bibnamefont{Smith}},
  \bibinfo{author}{\bibfnamefont{W.~J.}~\bibnamefont{Padilla}},
  \bibinfo{author}{\bibfnamefont{D.~C.} \bibnamefont{Vier}},
  \bibinfo{author}{\bibfnamefont{S.~C.} \bibnamefont{Nemat-Nasser}},
  \bibnamefont{and} \bibinfo{author}{\bibfnamefont{S.}~\bibnamefont{Schultz}},
  \bibinfo{journal}{Phys. Rev. Lett.} \textbf{\bibinfo{volume}{84}},
  \bibinfo{pages}{4184} (\bibinfo{year}{2000}).

\bibitem[{\citenamefont{Pendry et~al.}(1999)\citenamefont{Pendry, Holden,
  Robbins, and Stewart}}]{pendrySRR}
\bibinfo{author}{\bibfnamefont{J.}~\bibnamefont{Pendry}},
  \bibinfo{author}{\bibfnamefont{A.}~\bibnamefont{Holden}},
  \bibinfo{author}{\bibfnamefont{D.}~\bibnamefont{Robbins}}, \bibnamefont{and}
  \bibinfo{author}{\bibfnamefont{W.}~\bibnamefont{Stewart}},
  \bibinfo{journal}{IEEE Trans. Microwave Th. and Tech.}
  \textbf{\bibinfo{volume}{47}}, \bibinfo{pages}{2075 } (\bibinfo{year}{1999}).

\bibitem[{\citenamefont{Ziolkowski}(2004)}]{ENZpaper}
\bibinfo{author}{\bibfnamefont{R.~W.}~\bibnamefont{Ziolkowski}},
  \bibinfo{journal}{Phys. Rev. E} \textbf{\bibinfo{volume}{70}},
  \bibinfo{pages}{046608} (\bibinfo{year}{2004}).

\bibitem[{\citenamefont{Yuan et~al.}(2009)\citenamefont{Yuan, Popa, and
  Cummer}}]{zlossactive}
\bibinfo{author}{\bibfnamefont{Y.}~\bibnamefont{Yuan}},
  \bibinfo{author}{\bibfnamefont{B.-I.} \bibnamefont{Popa}}, \bibnamefont{and}
  \bibinfo{author}{\bibfnamefont{S.}~\bibnamefont{Cummer}},
  \bibinfo{journal}{Opt. Express} \textbf{\bibinfo{volume}{17}},
  \bibinfo{pages}{16135} (\bibinfo{year}{2009}).

\bibitem[{\citenamefont{Popa and Cummer}(2007)}]{bogdanactive}
\bibinfo{author}{\bibfnamefont{B.-I.} \bibnamefont{Popa}} \bibnamefont{and}
  \bibinfo{author}{\bibfnamefont{S.}~\bibnamefont{Cummer}},
  \bibinfo{journal}{Microwave and Optical Technology Letters}
  \textbf{\bibinfo{volume}{49}}, \bibinfo{pages}{2574} (\bibinfo{year}{2007}).

\bibitem[{\citenamefont{Fang et~al.}(2009)\citenamefont{Fang, Koschny, Wegener,
  and Soukoulis}}]{fang_PRBcalc}
\bibinfo{author}{\bibfnamefont{A.}~\bibnamefont{Fang}},
  \bibinfo{author}{\bibfnamefont{T.}~\bibnamefont{Koschny}},
  \bibinfo{author}{\bibfnamefont{M.}~\bibnamefont{Wegener}}, \bibnamefont{and}
  \bibinfo{author}{\bibfnamefont{C.~M.} \bibnamefont{Soukoulis}},
  \bibinfo{journal}{Phys. Rev. B} \textbf{\bibinfo{volume}{79}},
  \bibinfo{pages}{241104(R)} (\bibinfo{year}{2009}).

\bibitem[{\citenamefont{Chen et~al.}(2006)\citenamefont{Chen, Padilla, Zide,
  Gossard, Taylor, and Averitt1}}]{padilla_nature06}
\bibinfo{author}{\bibfnamefont{H.-T.} \bibnamefont{Chen}},
  \bibinfo{author}{\bibfnamefont{W.~J.} \bibnamefont{Padilla}},
  \bibinfo{author}{\bibfnamefont{J.~M.~O.} \bibnamefont{Zide}},
  \bibinfo{author}{\bibfnamefont{A.~C.} \bibnamefont{Gossard}},
  \bibinfo{author}{\bibfnamefont{A.~J.} \bibnamefont{Taylor}},
  \bibnamefont{and} \bibinfo{author}{\bibfnamefont{R.~D.}
  \bibnamefont{Averitt1}}, \bibinfo{journal}{Nature}
  \textbf{\bibinfo{volume}{444}}, \bibinfo{pages}{597} (\bibinfo{year}{2006}).

\bibitem[{\citenamefont{Reynet and Acher}(2004)}]{reynet_apl}
\bibinfo{author}{\bibfnamefont{O.}~\bibnamefont{Reynet}} \bibnamefont{and}
  \bibinfo{author}{\bibfnamefont{O.}~\bibnamefont{Acher}},
  \bibinfo{journal}{Applied Physics Letters} \textbf{\bibinfo{volume}{84}},
  \bibinfo{pages}{1198} (\bibinfo{year}{2004}).

\bibitem[{\citenamefont{Wang et~al.}(2007)\citenamefont{Wang, Ran, Chen, Mu,
  Kong, and Wu}}]{wang_apl07}
\bibinfo{author}{\bibfnamefont{D.}~\bibnamefont{Wang}},
  \bibinfo{author}{\bibfnamefont{L.}~\bibnamefont{Ran}},
  \bibinfo{author}{\bibfnamefont{H.}~\bibnamefont{Chen}},
  \bibinfo{author}{\bibfnamefont{M.}~\bibnamefont{Mu}},
  \bibinfo{author}{\bibfnamefont{J.~A.} \bibnamefont{Kong}}, \bibnamefont{and}
  \bibinfo{author}{\bibfnamefont{B.-I.} \bibnamefont{Wu}},
  \bibinfo{journal}{Applied Physics Letters} \textbf{\bibinfo{volume}{91}},
  \bibinfo{pages}{164101} (\bibinfo{year}{2007}).

\bibitem[{\citenamefont{Syms et~al.}(2008)\citenamefont{Syms, Solymar, and
  Young}}]{Syms_metamat2008}
\bibinfo{author}{\bibfnamefont{R.}~\bibnamefont{Syms}},
  \bibinfo{author}{\bibfnamefont{L.}~\bibnamefont{Solymar}}, \bibnamefont{and}
  \bibinfo{author}{\bibfnamefont{I.}~\bibnamefont{Young}},
  \bibinfo{journal}{Metamaterials} \textbf{\bibinfo{volume}{2}},
  \bibinfo{pages}{122 } (\bibinfo{year}{2008}).

\bibitem[{\citenamefont{Wang et~al.}(2008)\citenamefont{Wang, Zhou, Koschny,
  and Soukoulis}}]{costas_optexp08}
\bibinfo{author}{\bibfnamefont{B.}~\bibnamefont{Wang}},
  \bibinfo{author}{\bibfnamefont{J.}~\bibnamefont{Zhou}},
  \bibinfo{author}{\bibfnamefont{T.}~\bibnamefont{Koschny}}, \bibnamefont{and}
  \bibinfo{author}{\bibfnamefont{C.~M.} \bibnamefont{Soukoulis}},
  \bibinfo{journal}{Opt. Express} \textbf{\bibinfo{volume}{16}},
  \bibinfo{pages}{16058} (\bibinfo{year}{2008}).

\bibitem[{\citenamefont{Huang et~al.}(2010)\citenamefont{Huang, Poutrina, and
  Smith}}]{smithnonlinear}
\bibinfo{author}{\bibfnamefont{D.}~\bibnamefont{Huang}},
  \bibinfo{author}{\bibfnamefont{E.}~\bibnamefont{Poutrina}}, \bibnamefont{and}
  \bibinfo{author}{\bibfnamefont{D.}~\bibnamefont{Smith}},
  \bibinfo{journal}{Applied Physics Letters} \textbf{\bibinfo{volume}{96}},
  \bibinfo{pages}{104104} (\bibinfo{year}{2010}).

\bibitem[{\citenamefont{Shadrivov et~al.}(2006)\citenamefont{Shadrivov,
  Morrison, and Kivshar}}]{nonlinearSRRs}
\bibinfo{author}{\bibfnamefont{I.}~\bibnamefont{Shadrivov}},
  \bibinfo{author}{\bibfnamefont{S.}~\bibnamefont{Morrison}}, \bibnamefont{and}
  \bibinfo{author}{\bibfnamefont{Y.}~\bibnamefont{Kivshar}},
  \bibinfo{journal}{Opt. Express} \textbf{\bibinfo{volume}{14}},
  \bibinfo{pages}{9344} (\bibinfo{year}{2006}).

\bibitem[{\citenamefont{Wang et~al.}(2009)\citenamefont{Wang, Luo, Peng,
  Huangfu, Jiang, Wang, Chen, and Ran}}]{activenonlinear}
\bibinfo{author}{\bibfnamefont{Z.}~\bibnamefont{Wang}},
  \bibinfo{author}{\bibfnamefont{Y.}~\bibnamefont{Luo}},
  \bibinfo{author}{\bibfnamefont{L.}~\bibnamefont{Peng}},
  \bibinfo{author}{\bibfnamefont{J.}~\bibnamefont{Huangfu}},
  \bibinfo{author}{\bibfnamefont{T.}~\bibnamefont{Jiang}},
  \bibinfo{author}{\bibfnamefont{D.}~\bibnamefont{Wang}},
  \bibinfo{author}{\bibfnamefont{H.}~\bibnamefont{Chen}}, \bibnamefont{and}
  \bibinfo{author}{\bibfnamefont{L.}~\bibnamefont{Ran}},
  \bibinfo{journal}{Applied Physics Letters} \textbf{\bibinfo{volume}{94}},
  \bibinfo{pages}{134102} (\bibinfo{year}{2009}).

\bibitem[{\citenamefont{Klein et~al.}(2006)\citenamefont{Klein, Enkrich,
  Wegener, and Linden}}]{wegener_science06}
\bibinfo{author}{\bibfnamefont{M.~W.} \bibnamefont{Klein}},
  \bibinfo{author}{\bibfnamefont{C.}~\bibnamefont{Enkrich}},
  \bibinfo{author}{\bibfnamefont{M.}~\bibnamefont{Wegener}}, \bibnamefont{and}
  \bibinfo{author}{\bibfnamefont{S.}~\bibnamefont{Linden}},
  \bibinfo{journal}{Science} \textbf{\bibinfo{volume}{313}},
  \bibinfo{pages}{502} (\bibinfo{year}{2006}).

\bibitem[{\citenamefont{Veselago}(1968)}]{veselago}
\bibinfo{author}{\bibfnamefont{V.}~\bibnamefont{Veselago}},
  \bibinfo{journal}{Sov. Phys. USPEKHI} \textbf{\bibinfo{volume}{10}},
  \bibinfo{pages}{509} (\bibinfo{year}{1968}).

\bibitem[{\citenamefont{Pendry}(2008)}]{PendryTimereversal}
\bibinfo{author}{\bibfnamefont{J.~B.} \bibnamefont{Pendry}},
  \bibinfo{journal}{Science} \textbf{\bibinfo{volume}{322}},
  \bibinfo{pages}{71} (\bibinfo{year}{2008}).

\bibitem[{\citenamefont{Yariv}(1978)}]{PCYariv}
\bibinfo{author}{\bibfnamefont{A.}~\bibnamefont{Yariv}},
  \bibinfo{journal}{IEEE J. Quant. Elec.} \textbf{\bibinfo{volume}{QE-14}},
  \bibinfo{pages}{650} (\bibinfo{year}{1978}).

\bibitem[{\citenamefont{Chiu et~al.}(2006)\citenamefont{Chiu, Yum, Chang, Xue,
  and Chan}}]{ChiuRetroarray}
\bibinfo{author}{\bibfnamefont{L.}~\bibnamefont{Chiu}},
  \bibinfo{author}{\bibfnamefont{T.}~\bibnamefont{Yum}},
  \bibinfo{author}{\bibfnamefont{W.}~\bibnamefont{Chang}},
  \bibinfo{author}{\bibfnamefont{Q.}~\bibnamefont{Xue}}, \bibnamefont{and}
  \bibinfo{author}{\bibfnamefont{C.}~\bibnamefont{Chan}},
  \bibinfo{journal}{Microwave and Optical Technology Letters}
  \textbf{\bibinfo{volume}{48}}, \bibinfo{pages}{409} (\bibinfo{year}{2006}).

\bibitem[{\citenamefont{Pon}(1964)}]{PonRetroarray}
\bibinfo{author}{\bibfnamefont{C.}~\bibnamefont{Pon}}, \bibinfo{journal}{IEEE
  Trans. Ant. Prop.} \textbf{\bibinfo{volume}{12}}, \bibinfo{pages}{176}
  (\bibinfo{year}{1964}).

\bibitem[{\citenamefont{Fusco et~al.}(2010)\citenamefont{Fusco, Buchanan, and
  Malyuskin}}]{subwaveExp}
\bibinfo{author}{\bibfnamefont{V.~F.} \bibnamefont{Fusco}},
  \bibinfo{author}{\bibfnamefont{N.~B.} \bibnamefont{Buchanan}},
  \bibnamefont{and}
  \bibinfo{author}{\bibfnamefont{O.}~\bibnamefont{Malyuskin}},
  \bibinfo{journal}{IEEE Trans. Ant. Prop.} \textbf{\bibinfo{volume}{58}},
  \bibinfo{pages}{798} (\bibinfo{year}{2010}).

\bibitem[{\citenamefont{Malyuskin and Fusco}(2010)}]{IEEEphaseconj}
\bibinfo{author}{\bibfnamefont{O.}~\bibnamefont{Malyuskin}} \bibnamefont{and}
  \bibinfo{author}{\bibfnamefont{V.}~\bibnamefont{Fusco}},
  \bibinfo{journal}{IEEE Trans. Ant. Prop.} \textbf{\bibinfo{volume}{58}},
  \bibinfo{pages}{459} (\bibinfo{year}{2010}).

\bibitem[{\citenamefont{Landau and Lifshitz}(1976)}]{landau_mech}
\bibinfo{author}{\bibfnamefont{L.~D.} \bibnamefont{Landau}} \bibnamefont{and}
  \bibinfo{author}{\bibfnamefont{E.~M.} \bibnamefont{Lifshitz}},
  \emph{\bibinfo{title}{Mechanics}} (\bibinfo{publisher}{Pergamon Press},
  \bibinfo{address}{New York}, \bibinfo{year}{1976}), \bibinfo{edition}{3rd}
  ed.

\end{thebibliography}
\end{document}